\documentclass[showpacs,amsmath,amssymb,twocolumn, 10pt]{revtex4}
\usepackage{graphicx,color}
\usepackage{amsmath}
\usepackage[font=scriptsize]{caption}
\usepackage{subcaption}
\usepackage{float}
\usepackage{color}
\usepackage{amssymb}
\usepackage{bbm}

\usepackage{amsfonts}
\usepackage{bm}

\newcommand{\id}{\mbox{id}}
\newcommand{\ud}{\mathrm{d}}

\newcommand{{\Cd}}{{\mathbb{C}^d}}

\newcommand{\tr}{\mathrm{Tr}}

\def\oper{{\mathchoice{\rm 1\mskip-4mu l}{\rm 1\mskip-4mu l}
{\rm 1\mskip-4.5mu l}{\rm 1\mskip-5mu l}}}
\def\<{\langle}
\def\>{\rangle}

\newtheorem{Example}{Example}

\newcommand{\beq}{\begin{equation}}
\newcommand{\eeq}{\end{equation}}
\newcommand{\bear}{\begin{eqnarray}}
\newcommand{\ear}{\end{eqnarray}}
\newcommand{\bdm}{\begin{displaymath}}
\newcommand{\edm}{\end{displaymath}}

\newcommand{\suma}{\sum_{\alpha=0}^3}
\newcommand{\al}{\alpha}
\newcommand{\sig}{\sigma}
\newcommand{\La}{\Lambda}
\newcommand{\la}{\lambda}
\newcommand{\G}{\Gamma}
\newcommand{\g}{\gamma}
\newcommand{\f}{\phi}
\newcommand{\F}{\Phi}

\begin{document}
\title{\textbf{
Markovian semigroup from non-Markovian evolutions}}
\author{Filip A. Wudarski}
\affiliation{Quantum Research Group, School of Chemistry and Physics,
University of KwaZulu-Natal, Durban 4001, South Africa,
and National Institute for Theoretical Physics (NITheP), KwaZulu-Natal, South Africa}
\author{Dariusz Chru\'sci\'nski}
\affiliation{Institute of Physics, Faculty of Physics, Astronomy and Informatics \\ Nicolaus Copernicus University,
Grudzi{a}dzka 5/7, 87--100 Torun, Poland}


\begin{abstract}
It is shown that a convex combination of two non-Markovian evolutions may lead to Markovian semigroup.  This shows that convex combination of quantum evolutions displaying nontrivial memory effects may  result in a perfectly memoryless evolution.
\end{abstract}
\maketitle


{\em Introduction} -- A general quantum evolution is represented by a dynamical map $\Lambda(t)$, i.e. a family of completely positive and trace-preserving maps such that $\rho \rightarrow \rho(t) = \Lambda(t)\rho$, where $\rho(t)$ denotes the density operator at time `$t$' \cite{Breuer}. One usually assumes that $\Lambda(t)$ satisfies time-local master equation
\begin{equation}\label{}
  \frac{d}{dt} \Lambda(t) = \mathcal{L}(t) \Lambda(t) \ ,
\end{equation}
where the time-local generator $\mathcal{L}(t)$ has the following well-known form
\begin{eqnarray}\label{L}
  \mathcal{L}(t)\rho &=& -i[H(t),\rho]  \\ &+& \sum_\alpha \gamma_\alpha(t) \left( V_\alpha(t) \rho V_\alpha^\dagger(t) - \frac 12 \{ V^\dagger_\alpha(t) V_\alpha(t),\rho\} \right) , \nonumber
\end{eqnarray}
with the time dependent Hamiltonian $H(t)$ and time dependent dissipative part governed by time dependent rates $\gamma_\alpha(t)$ and noise operators $V_\alpha(t)$. Recently a lot of attention was devoted to the analysis of Markovianity of quantum evolution represented by $\Lambda(t)$ (see recent review papers \cite{rev1,rev2}). Let us recall that $\Lambda(t)$ is called divisible (or CP-divisible) if
\begin{equation}\label{}
  \Lambda(t) = V(t,s)\Lambda(s) ,
\end{equation}
and $V(t,s)$ is completely positive for all $t\geq s$ \cite{cubitt,RHP}. This property is fully characterized by the time-local generator: $\Lambda(t)$ is CP-divisible if and only if $\gamma_\alpha(t) \geq 0$ for all $t \geq 0$. One of the approaches to quantum Markovianity states that  quantum evolution
is Markovian iff the corresponding dynamical map  $\Lambda(t)$ is CP-divisible \cite{cubitt,RHP}. A slightly weaker notion of Markovianity was proposed in \cite{BLP}. The advantage of BLP approach \cite{BLP} is an operational characterization based on the following definition: $\Lambda(t)$ is Markovian if for any $\rho_1$ and $\rho_2$
\begin{equation}\label{blp}
  \frac{d}{dt} || \Lambda(t)[\rho_1-\rho_2]||_1 \leq 0 ,
\end{equation}
where $||X||_1 = {\rm Tr}\sqrt{X^\dagger X}$ denotes the trace norm of $X$. In this paper we attribute  Markovianity to the notion of CP-divisibility. However, the main example we use to illustrate the paper does not distinguish between these two notions.

\vspace{.2cm}

 {\em Convex combination of Markovian evolutions} --
Note that if $\mathcal{L}_1(t)$ and $\mathcal{L}_2(t)$ are Markovian generators then $\alpha_1 \mathcal{L}_1(t) + \alpha_2 \mathcal{L}_2(t)$ is again Markovian generator for arbitrary $\alpha_1,\alpha_2 \geq 0$. Hence Markovian generators define a convex set (actually a convex cone) in the space of all admissible time-local generators (\ref{L}). It is no longer true on the level of dynamical maps, i.e. if $\Lambda_1(t)$ and $\Lambda_2(t)$ are Markovian (i.e. CP-divisible), then $\alpha_1 \Lambda_1(t) + \alpha_2 \Lambda_2(t)$ needs not be CP-divisible \cite{cubitt}. A simple example illustrating that the space of CP-divisible maps is not convex was recently provided in \cite{fi3}: consider two Markovian semigroups generated by
\begin{equation}\label{}
  \mathcal{L}_1\rho = \frac c2 [ \sigma_1 \rho \sigma_1 - \rho] \ \ \ ;\  \mathcal{L}_2\rho = \frac c2 [ \sigma_2 \rho \sigma_2 - \rho]\ ,
\end{equation}
where $c > 0$ , and $\sigma_1,\sigma_2$ are Pauli matrices. One finds for the convex combinations $\Lambda(t)=\frac 12 [ e^{t\mathcal{L}_1} +  e^{t\mathcal{L}_2}]$
\begin{equation}\label{}
  \Lambda(t)\rho = \frac{1+e^{-ct}}{2}\, \rho + \frac{1-e^{-ct}}{4}\, (\sigma_1 \rho \sigma_1 + \sigma_2 \rho \sigma_2) \ .
\end{equation}
Clearly $\Lambda(t)$ is a legitimate dynamical map but it is not CP-divisible. Indeed, the corresponding time-local generator reads
\begin{equation}\label{}
  \mathcal{L}(t)\rho = \sum_{k=1}^3 \gamma_k(t) [ \sigma_k \rho \sigma_k - \rho] ,
\end{equation}
where
\begin{equation}\label{tanh}
  \gamma_1(t) = \gamma_2(t) = \frac c2 \ , \ \ \gamma_3(t)= - \frac c2 \tanh(ct)\ ,
\end{equation}
and evidently leads to non-Markovian evolution due to $\gamma_3(t) < 0$. This generator was analyzed in \cite{erika} as an example of {\em eternal} non-Markovianity.  This simple example shows that convex combination of Markovian evolutions (even semigroups!) might lead to legitimate non-Markovian evolution. Interestingly, this example may be easily generalized for qudit systems \cite{fi3}.

\vspace{.2cm}

{\em Convex combination of non-Markovian evolutions} --
In the present paper we provide a simple example showing that a convex combination of two non-Markovian evolution may lead to Markovian semigroup.
Consider a quantum channel $\mathcal{E}$ for a qudit system which defines a projector, that is, $\mathcal{E}^2 = \mathcal{E}$. A typical example is a channel which maps arbitrary state $\rho$ into a fixed state $\omega$, that is, $\mathcal{E}\rho = \omega {\rm Tr}\rho$. If $\omega =  \frac 1d \mathbb{I}$ then $\mathcal{E}$ is a  completely depolarizing channel. Taking an orthonormal basis $\{|1\>\,\ldots,|d\>\}$ in $\mathbb{C}^d$ one may define another CPTP projector via $\mathcal{E}\rho = \sum_{k=1}^d |k\>\<k| \rho|k\>\<k|$. Now, for arbitrary CPTP projector $\mathcal{E}$ let us consider the following Markovian generator
\begin{equation}\label{}
  \mathcal{L} = \mathcal{E} - \oper .
\end{equation}
We show that for a given $\gamma >0$ one can find time dependent $\gamma_1(t)$ and $\gamma_2(t)$ such that the following Markovian semigroup
$$ \Lambda(t) = e^{\gamma\mathcal{L}t}  = e^{-\gamma t} \oper + [1-e^{-\gamma t}] \mathcal{E} , $$
may be constructed as a convex combination
\begin{equation}\label{}
  \Lambda(t) = p \Lambda_1(t) + (1-p) \Lambda_2(t) ,
\end{equation}
with
\begin{equation}\label{}
  \Lambda_k(t) = \exp( \Gamma_k(t) \mathcal{L}) = e^{-\Gamma_k(t)} \oper + [1- e^{-\Gamma_k(t)}] \mathcal{E} ,
\end{equation}
and $\Gamma_k(t) = \int_0^t \gamma_k(\tau)d\tau$. Moreover, neither $\Lambda_1(t)$ nor $\Lambda_2(t)$ is Markovian which means that $\gamma_k(t) \ngeq 0$.

Note, that $\Lambda_k(t)$ is CP if and only if $\mu_k(t) = e^{-\Gamma_k(t)} \in [0,1]$.  One has the following relation
\begin{equation}\label{I}
   e^{-\gamma t} = p \mu_1(t) + (1-p) \mu_2(t)  . 
\end{equation}
Let $\mu_1(t) = \frac 1p e^{-\gamma t} g(t)$ and hence
$$   \mu_2(t) = e^{-\gamma t} \frac{1-g(t)}{1-p} . $$
The initial condition $\Lambda_k(0)=\id$ implies $\mu_1(0) = \mu_2(0) = 1$, hence  $g(0)=p$.  Moreover, the constraint $0 < \mu_1(t) \leq 1$ implies $0 < g(t) \leq p$. Now, to satisfy $0 < \mu_2(t) \leq 1$ let us consider the following $g(t)$: $g(t) = p$ for $t \in [0,t_*]$ and $p(t) \in [0,p]$ for $t > t_*$, where $t_*$ is defined via the relation
$$   e^{-\gamma t_*} = 1- p , $$
which implies $t_* = - \frac 1\gamma \ln(1-p)$. Finally,
\begin{equation}\label{}
  \mu_1(t) = \left\{ \begin{array}{ll} e^{-\gamma t} & \ \ t \in [0,t_*] \\
  e^{-\gamma t} \frac{g(t)}{p} & \ \ t > t_* \end{array} \right.
\end{equation}
and
\begin{equation}\label{}
  \mu_2(t) = \left\{ \begin{array}{ll} e^{-\gamma t} & \ \ t \in [0,t_*] \\
  e^{-\gamma t} \frac{1-g(t)}{1-p} & \ \ t > t_* \end{array} \right.
\end{equation}
The corresponding local depolarizing rates read
 $$   \gamma_1(t) = \gamma - \frac{\dot{g}(t)}{g(t)}  \ , \ \   \gamma_2(t) = \gamma + \frac{\dot{g}(t)}{1-g(t)} $$
and hence for $t \leq t_*$ one has $ \gamma_1(t) = \gamma_2(t) = \gamma$. As an example let us  consider
\begin{equation}\label{}
  g(t) = p \left\{ 1 - \varepsilon \sin^2(\gamma[t-t_*]) H(t-t_*) \right\} ,
\end{equation}
where $H(t)$ denotes the Heaviside step function, and $0 < \varepsilon < 1$. The corresponding local depolarizing rates for $\gamma=1,\varepsilon=\frac{3}{4}$ and $p=\frac{3}{4}$ are displayed in Fig. 1. It is evident that both $\gamma_1(t)$ and $\gamma_2(t)$ become negative for some time intervals and hence neither $\Lambda_1(t)$ nor $\Lambda_2(t)$ is CP-divisible. It is worth stressing that manipulation of parameters  $\gamma,\varepsilon$ and $p$ can gives us different types of behaviour, from CP-divisible for both $\Lambda_1(t)$ and $\Lambda_2(t)$ to non-divisible for either or both of $\Lambda_1(t)$ and $\Lambda_2(t)$.

\begin{figure}  \label{F1}
\includegraphics[width=6cm]{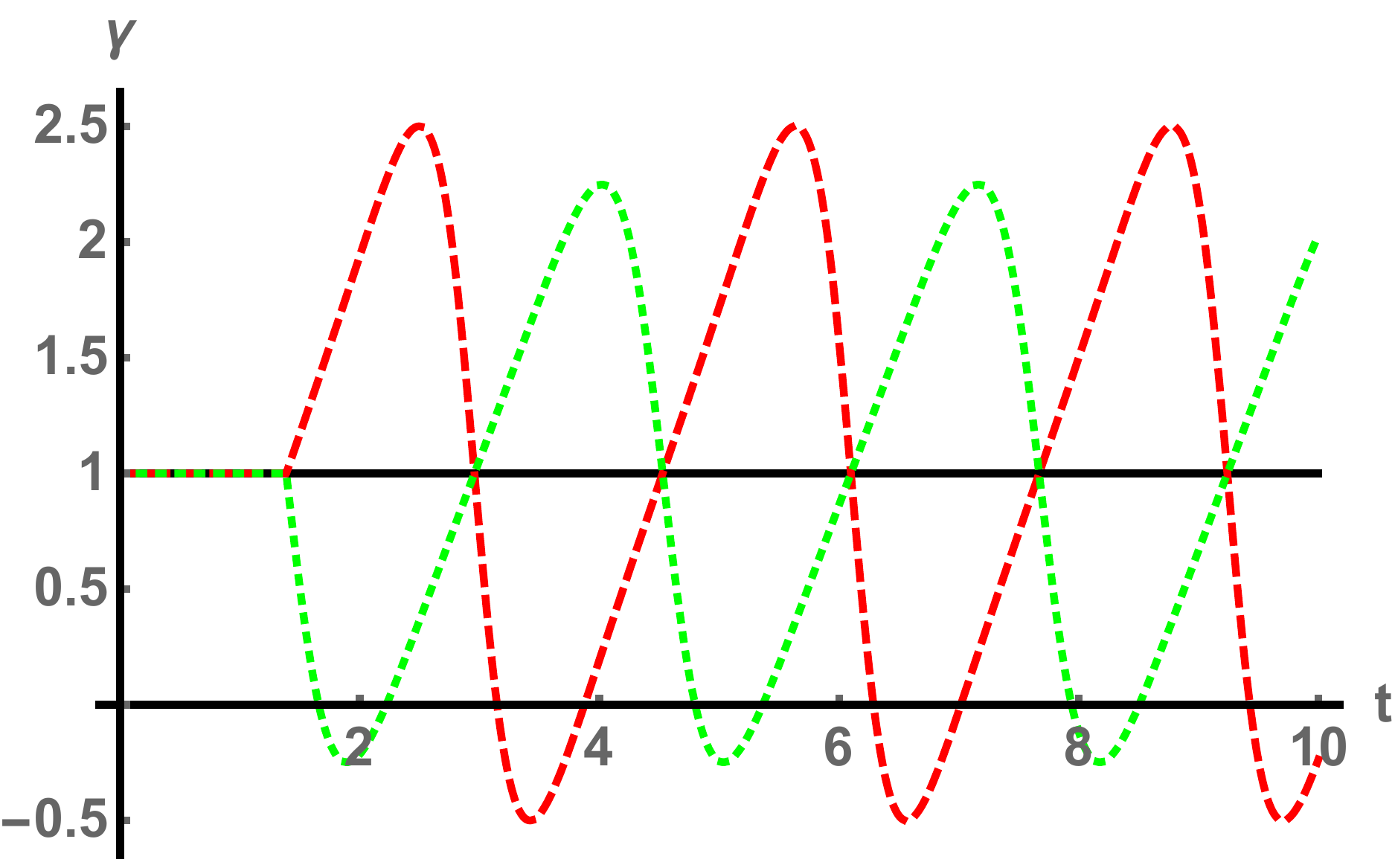}
\caption{Local decoherence rates for three types of dynamical maps with $\gamma=1,\varepsilon=\frac{3}{4}$ and $p=\frac{3}{4}$. The black line represents Markovian semigroup. The dashed (red) and the dotted  (green) are $\gamma_1(t)$ and $\gamma_2(t)$ respectively, and due to its interval negativity, they represent non-Markovian dynamics.}
\end{figure}

\begin{figure}   \label{F2}
\includegraphics[width=6cm]{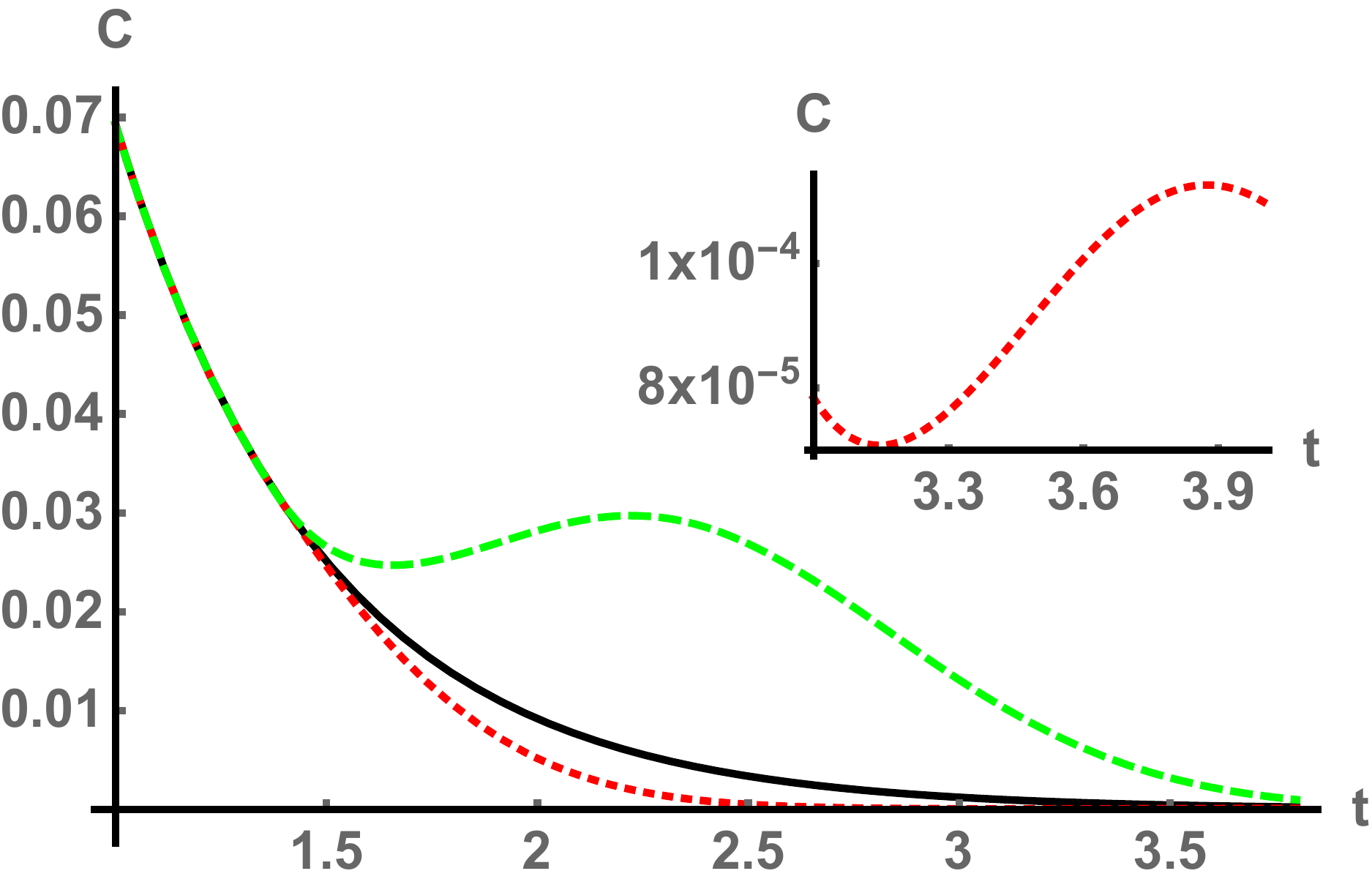}
\caption{Channel capacity $\mathcal{C}$ of three dynamical maps with $\gamma=1,\varepsilon=\frac{3}{4}$ and $p=\frac{3}{4}$. The black line corresponds to Markovian semigroup while the dashed (red) and the dotted (green) are channel capacities for non-Markovian dynamics $\Lambda_1(t)$ and $\Lambda_2(t)$ respectively.  In the green, we may clearly observe nonmonotonic behaviour. In the upper right corner we may observer nonmonotonicity of $\Lambda_2(t)$ for $t\in(3,4)$. Therefore, both of $\Lambda_1(t)$ and $\Lambda_2(t)$ are non-Markovian. }
\end{figure}

Non-Markovianiny of quantum evolution represented by $\Lambda(t)$ may be also analyzed in terms of channel capacity \cite{Bogna}.  If $\Lambda(t)$ is CP-divisible then
\begin{equation}\label{}
  \frac{d}{dt} \mathcal{C}(\Lambda(t)) \leq 0 \ ,
\end{equation}
i.e. capacity monotonically decreases.
For  depolarising channels one may easily  evaluate channel capacity \cite{cap}: for $\mathcal{E}_\lambda = \lambda \oper + (1-\lambda) \mathcal{E}$ one has
\begin{equation}\label{}
  \mathcal{C}(\mathcal{E}_\lambda) = \ln d - S_{\min}(\mathcal{E}_\lambda) ,
\end{equation}
where the minimal output entropy reads
\begin{eqnarray*}\label{}
  S_{\min}(\mathcal{E}_\lambda) &=& - \left(\lambda + \frac{1-\lambda}{d}\right) \ln  \left(\lambda + \frac{1-\lambda}{d}\right)\nonumber \\ & -& (d-1)  \frac{1-\lambda}{d}  \ln  \frac{1-\lambda}{d} .
\end{eqnarray*}
The corresponding plots of capacities in the qubit case are provided in Fig. 2. It is evident that $\Lambda_k(t)$, $k=1,2$ displays highly non-Markovian behaviour.

\vspace{.2cm}

{\em Semi-Markov evolution} -- Quantum evolution generated by the time-local generator $\mathcal{L}(t) = \gamma(t)[\mathcal{E} - \oper]$, with $\mathcal{E}$ being a CPTP projector may be equivalently described in terms of non-local memory kernel
\begin{equation}\label{}
  \mathcal{K}(t) = k(t)[\mathcal{E} - \oper] ,
\end{equation}
for some memory function $k(t)$ \cite{NJP-2011,EPL}. The corresponding non-local master equation
\begin{equation}\label{}
  \frac{d}{dt} \Lambda(t) =\int_0^t \mathcal{K}(t-\tau) \Lambda(\tau) d\tau ,
\end{equation}
gives rise to the following solution
\begin{equation}\label{f}
  \Lambda(t) = \left(1 - \int_0^t f(\tau) d\tau\right) \oper  + \int_0^t f(\tau) d\tau\, \mathcal{E} ,
\end{equation}
and the function $f(t)$ is related to the memory function $k(t)$ via
\begin{equation}\label{}
  \widetilde{k}(s) = \frac{s\widetilde{f}(s)}{1-\widetilde{f}(s)} ,
\end{equation}
where $\widetilde{f}(s) = \int_0^\infty e^{-st}f(t)dt$ denotes the corresponding Laplace transform. One calls $\Lambda(t)$ {\em semi-Markov} if $f(t) \geq 0$ and $\int_0^\infty f(\tau)d\tau \leq 1$. In this case $f(t)$ plays the role of so-called waiting time distribution and $1- \int_0^t f(\tau) d\tau$ is interpreted as so-called survival probability.  Interestingly, it is well known that in the class (\ref{f}) the evolution $\Lambda(t)$ is semi-Markovian if and only if it is Markovian (CP-divisible) \cite{NJP-2011}. Moreover, $\Lambda(t)$ defined by (\ref{f}) defines a semigroup iff $f(t) = \gamma e^{-\gamma t}$.
Our example shows that convex combination of two evolutions which are not semi-Markov, i.e. $f_1(t),f_2(t) \ngeq 0$, and hence non-Markovian, may results in Markovian semigroup:
\begin{equation}\label{II}
  p f_1(t) + (1-p) f_2(t) = \gamma e^{-\gamma t} ,
\end{equation}
for $t \geq 0$. Note, that  (\ref{II}) reproduces (\ref{I}), that is, $f_k(t) = \gamma \mu_k(t)$.

\vspace{.2cm}

{\em Conclusions} --- A set of Markovian (CP-divisible) evolutions is not convex. It is shown that a convex combination of two non-Markovian evolutions may lead to Markovian semigroup. Similarly, using memory kernel master equation we shown that convex combination of quantum evolutions which are not semi-Markov (and hence non-Markovian) may result in Markovian semigroup. This shows that convex combination of quantum evolutions displaying nontrivial memory effects may kill all  memory effects and result in a perfectly memoryless evolution.

\vspace{.2cm}

{\em Acknowledgements} --- D.C. was partially supported by the National Science Center project 2015/17/B/ST2/02026.

\end{document}